# Mobility, traffic and radio channel prediction: 5G and beyond applications

Henrik Rydén, Alex Palaios, László Hévizi, David Sandberg, Tor Kvernvik, Hamed Farhadi

Ericsson Research, Ericsson AB, Sweden
Contact email: henrik.a.ryden@ericsson.com

*Abstract*— Machine learning (ML) is an important component for enabling automation in Radio Access Networks (RANs). The work on applying ML for RAN has been under development for many years and is now also drawing attention in 3GPP and Open-RAN standardization fora. A key component of multiple features, also highlighted in the recent 3GPP specification work, is the use of mobility, traffic and radio channel prediction. These types of predictions form the intelligence enablers to leverage the potentials for ML for RAN, both for current and future wireless networks. This paper provides an overview with evaluation results of current applications that utilize such intelligence enablers, we then discuss how those enablers likely will be a cornerstone for emerging 6G use cases such as wireless energy transmission.

## I. Introduction

Machine learning (ML) as an important component for enabling automation in radio access networks (RANs) is growing more important each year due to the densification of networks and the growth of generated data in cellular networks. There are several use cases showing the benefit of applying ML in RAN, for example the authors in [1] discuss the use of ML to enable the network to adjust the base station's antenna array tilt to improve the network performance. Another example is the work in [2], that details how ML improves the performance of the receivers to cope with RF hardware impairments for millimeter Wave operation. In another publication it is shown how ML can be used as a bridge to enable smooth transition from 4G to 5G for improving the spectrum sharing among the two technologies [3].

Many of the existing successful proposals, in this area, utilize ML-based *traffic prediction* model such as the one proposed in [4]. Another application of ML is *mobility prediction*. The sequence of radio-fingerprints is used to predict the mobility of devices; the predicted mobility can for example be used to improve paging, scheduling and handover management [5]. Another prediction-based application is the use of *radio channel prediction*, which in essence is the ability to predict a set of radio signal qualities, based on a subset of radio signal measurements. In Figure 1, we refer a certain radio signal measurement to a "dimension", this hence comprises predicting one radio dimension based on measurements on other radio dimensions. One such example is when the radio strength on a secondary carrier frequency is predicted based on radio measurements of the primary carrier frequency. In that use case the device only performs inter-frequency measurements when it has high probability of being in the coverage of the secondary carrier [6].

Driven by the huge attention from academia and industry in applying ML for 5G, 3GPP has initiated a study item in Release-17 with the objective to study the artificial intelligence (AI) standardization impact in a use-case driven approach. For each use case, the study item objective includes understanding potential ML standardization impact on the node or function in current Next Generation RAN architecture. For example, the 3GPP study item [7] discusses the changes needed on the network interface(s) to convey the input data to an ML-based function. The ML-based function can further introduce the signaling of an output data in form of a predicted value with an associated confidence interval. The ongoing discussions further highlight the usage of traffic, mobility, and radio channel prediction. The application of mobility and traffic load forecasts are discussed in the considered use cases, i.e., mobility, energy saving and load-balancing [7]. Moreover, in the upcoming RAN1 study item in Rel-18, one of the agreed use cases is to study AI-enabled beam management. The use of radio channel prediction is likely a vital component, such as predicting a set of beams based on measurements on another set of beams.

When it comes to the cellular networks' evolution towards the 6G generation, use cases such as supporting zero-energy devices, sensing, and further enhanced mobile broadband are few of the many proposed use cases [8]. ML is one of the cornerstones for building such systems, and there are many aspects that need to be addressed to enable a successful implementation.

In this paper, we first provide a set of examples including results of how to predict mobility, traffic and radio channel conditions. Next, we discuss how such predictions lead to various network performance improvements. Finally, we take a forward-looking approach and discuss how these predictions are also likely to be a cornerstone for the ongoing work on 6G. We exemplify and discuss several 6G use cases that can be enhanced from mobility, traffic and radio channel predictions. It should be noted that there are many other ML-based components, that are not discussed here because of space limitations, of building an intelligent 6G system, such as channel compression, scheduling, link adaptation etc.

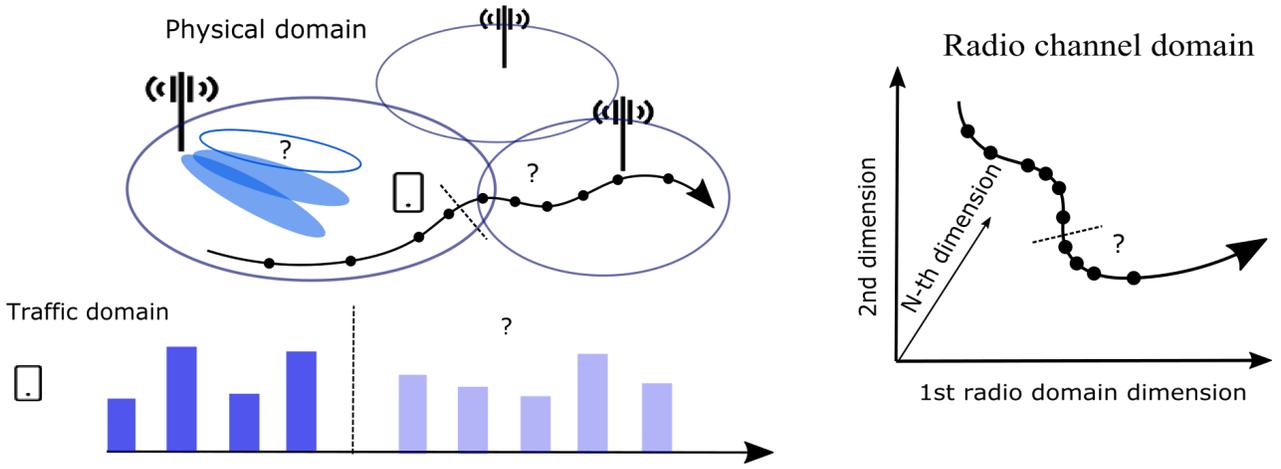

*Figure 1: Overview of mobility, traffic, and radio channel prediction. The radio channel/mobility prediction is shown both from a physical and radio channel domain perspecive. Radio channel prediction can comprise of predicting a non-measured beam (predict a certain radio dimension). Mobility prediction can comprise forecasting a certain radio signal quantity (in multiple radio dimensions). Traffic prediction is illustrated by showing an example of how packet sizes vary in time.*

**The paper is organized as follows**: In Section II we outline the various intelligence enablers; we continue in Section III where we showcase how such enablers translate into network performance improvement in current networks. We conclude in Section IV, where we provide an outlook into how such enablers can be used also for a future 6G system.

## II. INTELLIGENCE ENABLERS

This section provides an overview and evaluation results of the intelligence enablers (mobility, traffic and radio channel prediction). The overall illustration of the predictors outlined in the paper are illustrated in the Figure 1.

### A. Traffic prediction

A common way of realizing traffic prediction is to monitor arrival times, sizes and directions of individual packets in a network flow and use those to make a prediction regarding the remainder of the flow in terms of volume, duration, number of packets and packet inter-arrival time in the uplink/downlink directions. This is illustrated in Figure 1.

It can be noted that there is significant information about the characteristics of a network flow in the IP addresses and ports of the source and destination that a predictor can leverage to improve prediction performance, which has been investigated in *[4]*. This has the advantage that meaningful predictions can be made directly after receiving the first IP packet in the flow and successively improved as more information becomes available.

We train three classifiers to predict if the volume of a network flow will exceed a volume threshold (1 kB, 10 kB, 100 kb) using features extracted from the first IP packet, i.e. IP addresses and TCP ports. The classifiers are trained on an public dataset *[9]* and the *Receiver Operating Characteristics* (ROCs) for these classifiers are depicted in *Figure 2*. From this figure it can be seen that all three classifiers work well using only data from the first IP-packet in the flow. This makes them very suitable for downstream tasks that benefit from quick decisions, like steering of high volume users to high capacity cells (see section III.A).

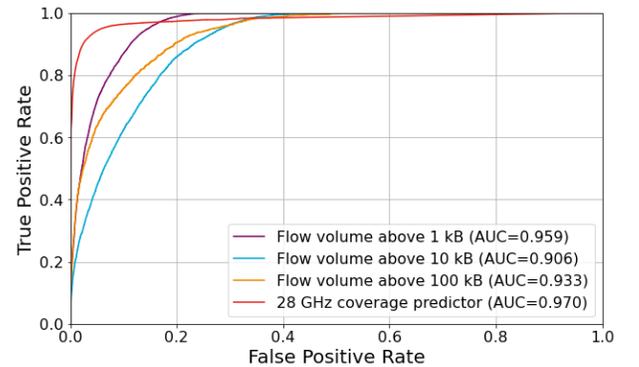

*Figure 2: Receiver operating characteristic (ROC) for the traffic volume predictor based on metadata from the first IP packet. ROC for the radio channel predictor, predicting 28 GHz coverage based on 3.5 GHz measurements*

### B. Radio channel prediction

In the area of radio channel prediction, the main idea is to measure on a subset of signal attributes and predict another set of signal attributes values. We define radio channel prediction as the instant or short-scale prediction in contrast to the mobility prediction, where a forecasted value in order of seconds is performed. The measured and predicted attributes could be in relation to the reference signals such as the 4G or 5G defined Channel State Information Reference Signal (CSI-RS), Synchronization Signal Block (SSB), etc. The signal attribute could be the signal quality quantity, for example Reference Signal Received Power (RSRP) defined in 5G.

One example of a radio channel predictor is, as aforementioned, to predict if having coverage on a secondary carrier frequency based on measurements on a primary carrier frequency. We evaluate such predictor using the setup described in [10], where a measurement comparing was conducted, measuring both 3.5 GHz and 28 GHz signal qualities for each location. Next, we divide the collected data into a training and test dataset and apply the random forest method to predict the coverage at 28 GHz frequency using the 3.5 GHz measurements. The performance of the radio channel predictor is shown by the ROC curve according to Figure 2. The Figure shows how one can achieve a ROC area under the curve (AUC) score of 0.970 which indicates that the selected random forest model can very accurately predict the 28 GHz coverage. In subsequent section, we further show how the predictor can be translated into improved traffic steering performance.

### C. Mobility prediction

The device mobility in the radio channel domain can be predicted using a sequence of radio signal measurements as shown in Figure 1. Device agnostic localization techniques are preferred, as they do not require localization information external to the cellular network. Both uplink and downlink radio measurements could be processed when predicting the mobility. Uplink measurements have lower data collection overhead since they do not require radio resource control (RRC) messaging on the air. In contrast, downlink measurements require RRC messaging both at the device and network side, however they have higher localization value compared to uplink measurements due to the multiple connection legs that device can measure on.

Among the mobility measurements defined in [11] the serving and neighbor-cell RSRPs have the most valuable localization information, and when coupled with timing-based measurements, it allows trilateration type of localization [12]. Advanced antenna systems with the capability of detecting the direction of arrival (DoA) of signals enhance further the previously mentioned localization techniques. Similarly, beamforming capable antennas, especially on higher frequencies, provide the direction of radio-wave propagation, hence permit localization of devices with sub-cell resolution.

One resource efficient positioning infrastructure is to only rely on the always available mobility measurements, instead of configuring extra radio resources associated to the 4G/5G defined positioning procedures. When mobility measurements are augmented with several other downlink properties measured and reported by active devices, such as strongest beam indices, then all those can be fed into an AI-based radio fingerprinting localization unit, which finds the location of the device in the radio domain shown in Figure 1. Each dimension of the radio domain is a radio measurement-based quantity, e.g. the RSRP from a cell identified by a physical cell ID. When expressed inside this radio domain a moving device defines a trajectory in this coordinate system.

Two kinds of learnings are possible based on the mobility prediction:
*1. Coverage prediction*: the large number of unanimous device measurements reveal the most common device trajectories in a region and the multi-cell, multi-carrier and multi-operator coverage along those trajectories. If the device trajectory follows a known pattern, then coverage prediction is possible for the device.

*2. Radio environment dynamics prediction*: By analyzing the recent history of device-related radio measurements, the dynamics of how the radio environment is changing for the device can be inferred. Radio resource allocations on the air and processing resources on the network side can take advantage of such inference, e.g. whether the device is static, moves slow or fast in the given heterogeneous network coverage. In terms of RRC, the network side is not interested in physical speed, rather in the dynamics of field variations.

Since devices often move along similar trajectories in an area, upon recognizing that a device follows a known mobility pattern, network management can take proactive, device-tailored RRC actions.

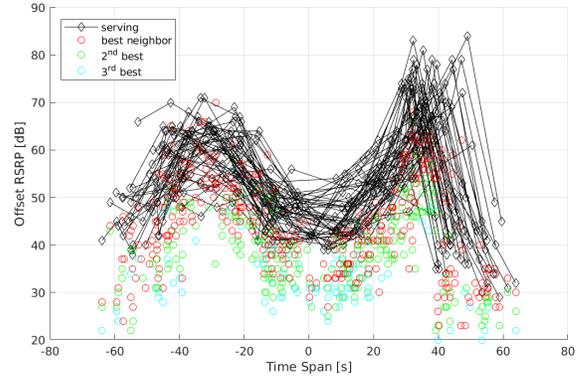

*Figure 3: Serving RSRP traces for the group of devices passing along an ML-mined trajectory*

As an example, Figure 3 shows the serving and neighbor-cell measurements from devices passing along a frequent trajectory, most probably along a highway segment which was part of the measurement campaign. Periodic neighbor cell measurement reports, in this case in every 10s in a 2 GHz LTE operator network, allowed us to localize devices and describe trajectories with sub-cell resolution and the shown RSRP traces have been seen along the picked trajectory.

### III. EXISTING APPLICATIONS OF INTELLIGENCE ENABLERS

In this section we focus on providing some examples of the benefits that ML-based predictions bring to the network. In this section, we exemplify how the use of the intelligence enablers in previous section can be translated into network performance improvement. The overview of the two exemplified existing applications is shown in Figure 4.

### A. Traffic steering use-case

To quantify how the traffic and radio channel predictors in previous sections can be used to improve network performance, we conduct an experiment where we assume that we have 100 devices connected to a primary cell operating at 3.5 GHz with partial coverage (22%) of a secondary cell at 28 GHz. Each device is associated with a traffic flow sampled from the dataset in [9]. In this dataset 12.5% of the flows have a total traffic volume that exceeds 10 kB. Next, we want to select a set of devices with coverage on the secondary cell and reasonably large traffic volume to be handed over to the secondary cell. For this we use the traffic predictor outlined in section II.

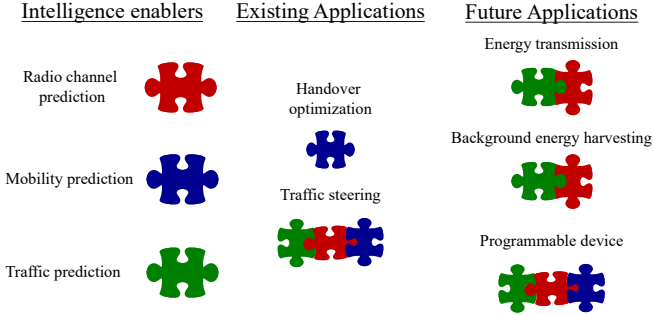

*Figure 4: Overview of intelligence enablers and the application to existing and future applications*

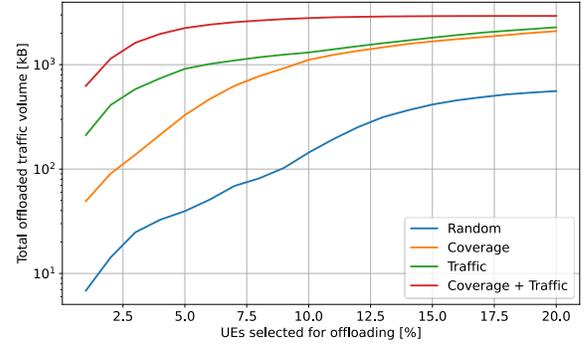

*Figure 5: Fraction of devices in one cell are selected for offloading based on the predicted probability that the total traffic volume will exceed 10 kB and that it will have coverage on the secondary carrier. The y-axis shows the total traffic volume for the devices selected for offloading.*

In Figure 5 we show the performance of four strategies for selecting which devices to hand over to the secondary carrier. We show the accumulated traffic volume that is offloaded by selecting the set of devices to be offloaded according to four strategies. Devices that are selected for offloading but is not covered by the secondary carrier does not contribute to the total offloaded traffic volume. For the *Random* strategy, devices are selected randomly with respect to both the coverage on the secondary carrier and the traffic volume. For the *Coverage* strategy, devices are selected with decreasing predicted probability of having coverage on the secondary (28 GHz) carrier but randomly with respect to the device traffic volume. For the *Traffic* strategy, devices are selected based on decreasing predicted probability of having a traffic volume that exceeds 10 kB but randomly with respect to coverage. Lastly, for the *Coverage+Traffic* strategy, predictions are combined and devices are selected based on the predicted probability of having coverage on the secondary carrier *and* a traffic volume that exceeds 10 kB. It should however be noted that with this way of viewing the results, detrimental effects such as handing over a device with poor coverage on the secondary cell or with very little traffic volume is not visible. But it can be seen that by cleverly selecting the set of devices to offload the effect of the offloading (i.e. the amount of traffic volume that is offloaded) can be increased by two orders of magnitude (compared to random selection) when relatively few devices are handed over.

In terms of network performance, the number of unnecessary measurements can be reduced from 78 % when performing random selection of devices, to a few percentages depending on how many devices to be handed over.

### B. Mobiliy-prediction: handover optimization

An illustrative example for the application of AI in RRC is the configuration management of the conditional handover. Conditional handover is a novel RRC procedure standardized by 3GPP [11] to make handovers prompt and robust. The main addition is that a potential target cell is prepared for the handover ahead of time. When the handover procedure is finalized, the target cell is ready to communicate with the new device entering the cell.

The conditional handover involves one or more potential target cells that need to allocate resources for the device before the handover condition is fulfilled and the RRC reconfiguration of the device with the connection information on those cells. The task of a ML-based algorithm in this procedure is to identify which cells are the candidate target cells and predict when the handover will occur.

ML is not only able to predict handover events for the devices, but also to customize the RRC configurations, including the parameters of handovers or measurement reporting, according to the instant mobility situation of devices. For example, by knowing the frequent trajectory patterns in a region and the mobility events that happen to devices along those trajectories, the network or cell-scope handover margin and time to trigger parameters can be tailored for the instant mobility of devices proceeding along a known trajectory.

### IV. INTELLIGENCE ENABLERS: TOWARDS 6G EVOLUTION

In this section we outline how the intelligence enablers can further support the evolution towards the 6G systems.

### A. Zero-energy devices – AI-enabled energy transfer

Wireless energy transfer is envisioned to provide energy to a wide range of devices including low-power and/or low-cost Internet of Things (IoT) devices, such as medical IoT devices [13] in the next generations of cellular networks. The widely spread network infrastructure in addition to data transmission can be used for transferring energy to low power devices such as IoT devices, where the devices harvest part or all their required energy from the transmitted radio frequency signals. Several techniques have been developed for simultaneous wireless information and power transfer (SWIPT) towards multiple devices, e.g., the one in [14].

There are several aspects that can degrade wireless energy transfer performance. For example, it is not a priori known which device, and when it requires energy [15]. Another problem is the prioritization of energy transfer over data transmission. Obtaining this information and adapting the energy transfer technique, accordingly, consumes energy, and thus reduces the system efficiency. A device might be idle for a long time and need to receive more energy to be able to receive a large packet within a certain time budget. It might also be a problem in case multiple devices need to receive energy at the same time-frequency resources.

Several prediction techniques such as *traffic prediction*, *radio energy prediction*, and *device energy prediction* can be applied as outlined in the following to enhance wireless energy transfer. One can utilize the *traffic prediction* outlined in this paper to improve the energy transmission toward devices. For example, first, a base station forecasts the amount of traffic which a device will transmit. Next, the base station adjusts the level of energy to be transmitted towards the intended device to provide the device with sufficient energy for the transmission of the predicted traffic. The overview of the idea is illustrated in Figure 6. Also, *radio-based energy prediction* could be used to predict the level of energy that a device can harvest in a certain time-frequency resource. In addition to radio channel and traffic prediction, one could also envision *device energy prediction*, that predicts the probability of a device not having the energy to receive nor transmit data. The inputs to the model could be device related data such as historical energy information data from device, the chipset energy consumption, or historical amount of energy transmitted from multiple base stations.

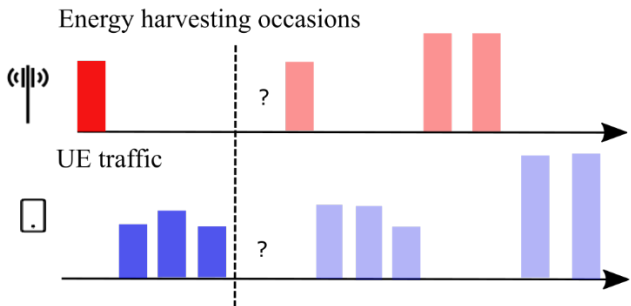

*Figure 6: Energy harvesting occasions based on traffic prediction*

### B. Background-Energy harvesting boosted using ML.

One of the key forces in shaping a 6G system is sustainability, where the network should be designed to limit the unnecessary energy expenditure. However, enabling zero-energy devices by energy harvesting at the device as proposed in the previous section, e.g. in the case of SWIPT systems, the dedicated transmission of energy to the device can lead to a high energy consumption overhead at the base station. In the traffic steering use case in III.A, we used the predicted coverage to find the best devices to offload from the network based on the radio and the use of forecasted traffic in the same offloading use case. One can apply the same idea for the energy harvesting use case, where prediction of traffic and radio can be used to find the carrier frequency which has the most available background energy (e.g., in the form of radio-frequency (RF) electromagnetic energy) for the device to harvest. Using background energy, i.e., energy intended for the data reception/transmission for another device, one can reduce the overall transmitted dedicated energy for the harvesting device.

Due to the network densification and deployment of a vast number of carriers, it is challenging to find what is the best resource for a device in terms of background energy to harvest. One can use the prediction of traffic and radio to obtain a background energy harvesting estimate for each time-frequency resource. This can be performed during the selection and the configuration of a device for performing energy harvesting in a given time-frequency resource. Using traffic and radio channel prediction, one can find the optimal time-frequency resource for a device to perform energy harvesting. Optimal in the sense of maximizing the energy harvested by the device without network having to spend dedicated energy transmission towards the device.

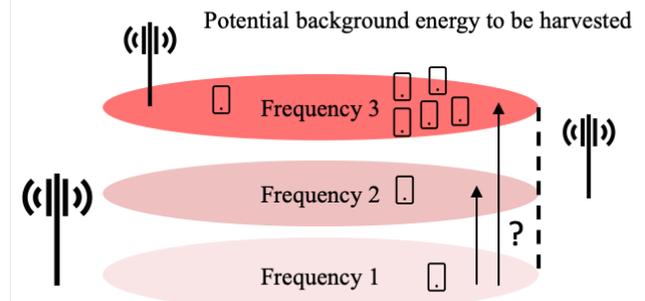

*Figure 7: Device inter-frequency handovers based on the amount of background energy that could be harvested by the device*

Note that one can use RL type of techniques to further optimize the selection of carrier frequency for energy harvesting.

### C. Efficient programmable environment via traffic and mobility predictions

One of the technologies that draws high attention from academia is a physical layer that is highly trainable, and thus adaptable to different environments. For example, in [16], the authors describe how such a system is possible to replace parts of a transceiver/receiver block with ML. It is not clear how such ML-based solution would be standardized. One option is to preconfigure a set of models, trained in an offline manner. This might hinder the ability to adapt to different environments. Another option would be to enable each, or group of base stations, to train their own ML-modules for a part of the air-interface, this would allow the model to adapt to the local deployment environment. These types of *programmable environments* are one of the technology visions for a 6G platform [8]. One drawback is that the network needs spending resources in downloading the model to the device, there is thus a cost associated to the configuring step.

To reduce such overhead cost, one can use traffic prediction to determine when one or more of the receiver chain blocks should be replaced by a ML-model. For a device with large data traffic volume the overhead in sending the model might be negligible. The use of traffic prediction can determine whether it is useful to download the model as illustrated in Figure 8, for example based on the previous device traffic or information from the service provider. Also, the mobility prediction can be used to configure one or more model(s) at the device, to configure models that are valid over the cells within the device predicted trajectory.

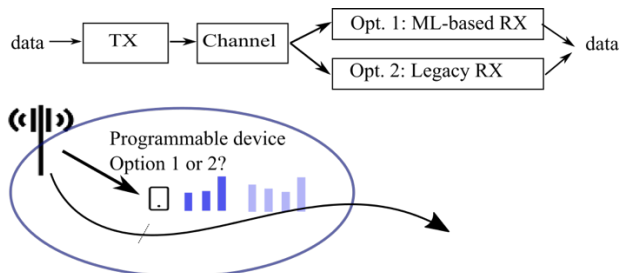

*Figure 8: Selecting ML-model for programmable device based on predicted traffic volume and mobility*

## V. CONCLUSION

The paper highlights the use of mobility, traffic and radio channel predictions as intelligence enablers. The network itself can further optimize its procedures and capabilities as evolving towards the 6G generation. Such new added intelligence to the network can further support more diverse use-cases and provide better experience to the end users. We have presented a vision on intelligence enablers as we think that ML has an important role to play in the transformation of next generations of cellular networks.